
\input phyzzx
%
%
%
\overfullrule=0pt
\sequentialequations
\def\({[}
\def\){]}
\def\lsquare{\left[}

%
%
\catcode`\@=11 

\def\myeqname#1#2{\rel@x {\pr@tect
  \ifnum\equanumber<0 \xdef#1{{\rm(\number-\equanumber#2)}}%
     \gl@bal\advance\equanumber by -1
  \else \gl@bal\advance\equanumber by 1
     \ifx\chapterlabel\rel@x \def\d@t{}\else \def\d@t{.}\fi
    \xdef#1{{\rm(\chapterlabel\d@t\number\equanumber#2)}}\fi #1}}

\def\eqreset{\rel@x {\pr@tect
  \ifnum\equanumber<0    \gl@bal\advance\equanumber by 1
  \else \gl@bal\advance\equanumber by -1\fi}}

\catcode`\@=12 


\def\bbold#1{\setbox0=\hbox{$#1$}%
     \kern-.04em\copy0\kern-\wd0
     \kern.08em\copy0\kern-\wd0
     \kern-.04em\box0 }

\def\Toprel#1\over#2{\mathrel{\mathop{#2}\limits^{#1}}}
\def\Botrel#1\over#2{\mathrel{\mathop{#1}\limits_{#2}}}

\def\slashb#1{\setbox0=\hbox{$#1$}#1\hskip-\wd0\dimen0=5pt\advance
       \dimen0 by-\ht0\advance\dimen0 by\dp0\lower0.5\dimen0\hbox
         to\wd0{\hss\sl/\/\hss}}
\def\Dslash{\slashb{D}}
\def\LL{{\cal L}}
\def\psibar{\bar\psi}

\def\alphaprime{\hbox{$(\alpha^{\prime})^{-1}$}}

\def\diag{\hbox{diag}}
\def\ds{\displaystyle}

\def\etal{\hbox{\it et al.}}

\def\mtilde{{\hbox{$\tilde m$}}}
\def\myfoot#1{\attach#1\vfootnote{#1}}

\def\pp{\phantom{-}}

\def\Tr{\,\hbox{Tr}\,}

\def\eff{\hbox{\tenpoint \it eff}\,}
\def\SeffU{S_{\kern -0.1em \eff}[u]}
\def\dminus{\partial_{-}}

\def\jprime{{j^{\prime}}}

\def\mprime{{m^{\prime}}}

\def\ac{{\alpha_c}}
\def\aj{{\alpha j}}

\def\al{{\alpha l}}

\def\bl{{\beta l}}

\def\alphaprime{{\alpha^{\prime}}}

\def\chiaj{{\chi_{}}_\aj}

\def\chial{{\chi_{}}_\al}

\def\chibl{{\chi_{}}_\bl}

\def\chip{\chi^{\prime}}
\def\chipaj{{\chip_{}}_\aj}

\def\naj{n_\aj}
\def\nal{n_\al}
\def\spi{\sqrt{4 \pi}}
\def\sspi{\sqrt{\pi}}

\def\PL{Phys. Lett.}

\REF\NQM{For a review see:
J. J. J. Kokkedee, {\sl The Quark Model},
W. A. Benjamin, New York, 1969.}
\REF\CA{For a detailed exposition of current algebra see:
\nextline
S. L. Adler and R.F. Dashen,
{\sl  Current Algebras and Applications to Particle Physics},
W.~A.~Benjamin, New York, 1968;\nextline
S.B. Treiman, R. Jackiw and D.J. Gross,
{\sl Lectures on Current Algebra and its Applications},
Princeton Univ. Press, 1972;\nextline
S.B. Treiman, R. Jackiw, B. Zumino and E. Witten,
{\sl Current Algebra and Anomalies},
Princeton Univ. Press, 1985.}

 \REF\GOR{M. Gell-Mann, R. J. Oakes and B. Renner \journal Phys. Rev.
 &175(68)2195 and references therein.}

\REF\GM{H. Georgi and A. Manohar \journal Nucl. Phys. &B310(88)527.}

\REF\KaplanI{
D. B. Kaplan \journal Phys. Lett. &B235(90)163;
{\sl Nucl. Phys.} {\bf B351}(1991), 137.}

\REF\Weinberg{S. Weinberg  \journal Phys. Rev. Lett. &65(90)1181.}

\REF\StechNRQM{U. Ellwanger and  B. Stech
\journal Z. Phys. &C49(91)683.}

\REF\EFK{J. Ellis, Y. Frishman and M. Karliner
\journal \PL &B272(91)333.}

\REF{\Skyrme}{T.H.R. Skyrme \journal Proc. Roy. Soc. London
&A260(1961)127.}

\REF{\SM}{E. Witten \journal   Nucl. Phys. &B223(1983)422,
{\sl ibid}  433;
G. Adkins, C. Nappi and \hbox{E. Witten}
 \journal Nucl. Phys. &B228(1983)433;
for the 3 flavour extension of the model see:
E. Guadagnini \journal Nucl. Phys. &236(1984)35;
P. O. Mazur, M.~A.~Nowak and \hbox{M. Prasza\l owicz},
 \sl Phys. Lett. \rm {\bf147B}(1984), 137.}

\REF\WittenLewes{E. Witten in Lewes Workshop Proc.; A. Chodos
{\sl et al.}, Eds; Singapore, World Scientific, 1984.}

\REF\tH{G. `t Hooft \journal Nucl. Phys. &B72(74)461;
{\bf B75}(1974), 461.}

\REF\CCG{C. G. Callan, N. Coote and D. J. Gross
\journal Phys. Rev. &D13(76)1649.}

\REF\Einhorn{M. Einhorn
\journal Phys. Rev. &D14(76)3451.}

\REF\BESW{R.C. Brower, J. Ellis, M.G. Schmidt and J.H. Weis
\journal Nucl. Phys. &B128(77)131
\journal {} &B128(77)175.}

\REF{\WittenBoson}{E. Witten \journal Comm. Math. Phys. &92(1984)455.}

\REF{\Gonzales}{D. Gonzales and A.N. Redlich
\journal Nucl. Phys. &B256(1985)621.}

\REF{\DFS}{G.D. Date, Y. Frishman and J. Sonnenschein,
{\sl Nucl. Phys.} {\bf B283}(1987), 365.}

\REF{\FS}{Y. Frishman and J. Sonnenschein,
{\sl Nucl. Phys.} {\bf B294}(1987), 801.}

\REF{\FZ}{Y.~Frishman and W.~J.~Zakrzewski
\journal Nucl. Phys. &B331(1990)781 and in {\sl Proc.
of the 25-th Int. High Energy Physics Conference}, Singapore(1990).}
\REF\FK{Y. Frishman and M. Karliner,
{\sl Nucl. Phys.} {\bf B334}(1990), 339.}
\REF\HBP{K.~Hornbostel, S.~J.~Brodsky and H.C.~Pauli
\journal Phys. Rev. &D41(1990)3814.}
\REF\Coleman{S. Coleman \journal Ann. Phys. &101(76)239.}
\REF{\Gepner}{D. Gepner
\journal Nucl. Phys. &B252(1985)481.}

\REF\COLSYS{U. Ascher, J. Christiansen and R.D. Russel
\journal ACM Trans. Math. Soft. &7(81)209.}

\REF\WI{N. Isgur and M.B. Wise \journal Phys. Lett. &B232(1989)113;
{\bf B237}(1990), 527.}

\REF\VS{M.B. Voloshin and M.A. Shifman \journal Yad. Fiz. &45
(1987)463 ({\sl Sov. J. Nucl. Phys.} {\bf47}(1988), 511).}

\REF\PolitzerWise{H.D. Politzer and M.B. Wise
\journal \PL &B206(88)68, \journal{} &B208(88)504.}

\REF\NW{S. Nussinov and W. Wetzel \journal Phys. Rev. &D36(87)130.}

\REF\FGG{A. Falk, H. Georgi, Grinstein and M.B.~Wise\journal
Nucl. Phys. &B343(1990)1.}

\REF\GrinsteinRev{For recent reviews, see
B. Grinstein,
{\sl Lectures on Heavy Quark Effective Theory},
SSCL-PREPRINT-17, Dec 1991,
to be published in Workshop on High Energy Phenomenology, Mexico City
Mexico, Jul 1-10, 1991;
H.~Georgi, {\sl Heavy Quark Effective Field Theory},
HUTP-91-A039; M.B.~Wise, {\sl New Symmetries of Strong Interactions},
Lake Louise Lectures, CALT-68-1721, Feb.~1991.}

\REF\Rho{
M.~Rho, D.O.~Riska and
N.N.~Scoccola \journal \PL &B251(90)597.}
\REF{\DonoNappi}{J. Donoghue and C. Nappi
\journal Phys. Lett.  &B168(1986)105;
J. Donoghue, in Proc. of II-nd Int. Conf.
on $\pi N$ Physics.}

\REF{\YabuKK}{H. Yabu \journal Phys. Lett. &B218(89)124;
D. B. Kaplan and I. Klebanov
\journal Nucl. Phys. &B335(90)45; for a review see
I. Klebanov in Proc.
NATO ASI on Hadron and Hadronic Matter, Cargese, 1989.}

\REF{\Weigel}{H. Weigel, J. Schechter, N.W. Park and U.-G. Mei\ss ner,
{\sl Phys. Rev.} {\bf D42}\break(1990)3177.}

\REF\GrinsteinMende{B. Grinstein and P.F. Mende, {\sl Heavy Mesons in Two
Dimensions}, SSCL-Preprint-64/BROWN HET-851, submitted
to {\sl Phys. Rev. Lett.}}

\REF\EGK{J. Ellis, E. Gabathuler and
M.~Karliner  \journal  Phys. Lett. &B217(88)173;
for a recent review of the evidence for this, see
R. Decker, M. Nowakowski and U. Wiedner,
CERN preprint PPE 92-010 (1992). }

\REF\DOZItheory{J.~F.~Donoghue
\journal Phys. Rev. &D33(86)1516;
%
M.~Bauer, B.~Stech and M.~Wirbel
\journal Z. Phys &C34(1987)103.}

\REF\DOZIdata{CLEO Collaboration (C.~Bebek \etal)
\journal  Phys. Rev. Lett. &56(1986)1893;
\hbox{ARGUS} Collaboration (H. Albrecht, \etal)
\journal Z. Phys. &C33(1987)359;
ACCMOR Collaboration (S.~Barlag \etal)
\journal \PL &B232(89)561.}

\REF\Anjos{Tagged Photon Spectrometer Collaboration
(J.C. Anjos \etal),
{\sl Study of the doubly Cabibbo suppressed decay
$D^+ \rightarrow  \phi K^+$ and the
singly Cabibbo suppressed decay
$D_s^+ \rightarrow  \phi K^+$},
 FERMILAB-PUB-91-331.}

\REF\Adamovich{WA82 Collaboration (M. Adamovich \etal),
{\sl Measurement of Relative Branching Fractions of $D^0$ Cabibbo
Suppressed Decays}, CERN preprint PPE-92-015 and references therein.}

%

\frontpagetrue
\null
\vskip -2cm
{\baselineskip 13pt
\null
\line{\hfill CERN-TH-6426/92}
\line{\hfill WIS-92/21-PH MAR}
\line{\hfill TAUP-1946-92}
\line{\hfill hep-ph@xxx/9204213}

\vskip 0.9cm
\centerline{\bf \fourteenpoint Quark solitons as constituents
 of hadrons}
\vskip 0.6cm
\centerline{\bf John Ellis\myfoot{a}{\rm e-mail: JOHNE@CERNVM.BITNET}}
\centerline{CERN -- Geneva}
\vskip .4cm
\centerline{\bf Yitzhak Frishman\myfoot{b}{\rm e-mail:
FNFRISHM@WEIZMANN.BITNET}
 and Amihay Hanany\myfoot{c}{\rm e-mail: FTAMI@WEIZMANN.BITNET}}
\centerline {Department of Physics, Weizmann Institute of Science }
\centerline {76100 Rehovot, Israel}
\vskip .4cm
\centerline{and}
\vskip .4cm
\centerline{\bf Marek Karliner\myfoot{d}{\rm e-mail:
MAREK@TAUNIVM.BITNET}}
\centerline{Raymond and Beverly Sackler Faculty of Exact Sciences}
\centerline{School of Physics and Astronomy}
\centerline{Tel-Aviv University, 69978 Tel-Aviv, Israel.}
\vskip 0.5cm

\abstract

We exhibit static  solutions of multi-flavour  QCD in two
dimensions that have  the quantum numbers of  baryons and mesons,
constructed out  of quark and anti-quark  solitons.  In isolation
the latter  solitons have  infinite energy, corresponding  to the
presence of a string carrying  the non-singlet colour flux off to
spatial infinity.
When  $N_c$  solitons  of  this  type  are  combined,  a  static,
finite-energy, colour  singlet solution is  formed, corresponding
to a baryon. Similarly, static  meson solutions are formed out of
a soliton and  an anti-soliton of different  flavours.  The stability
of the mesons against annihilation is ensured by  flavour conservation.
The static solutions  exist only when the  fundamental fields of
the  bosonized Lagrangian  belong  to $U(N_c{\times}N_f)$  rather
than to $SU(N_c) \times U(N_f)$.
Discussion of flavour symmetry breaking requires a  careful treatment of
the normal ordering ambiguity.
Our  results can  be viewed  as  a  derivation of  the
constituent quark model in QCD$_2$,  allowing a detailed study of
constituent  mass generation and of the heavy quark symmetry.

\vskip 0.5cm
\vfill
\line{CERN-TH-6426/92 \hfill}
\line{March 1992 \hfill}
} 
\endpage
\pagenumber=1


\chapter{Introduction}

One of the key outstanding problems in strong interaction physics is the
derivation of hadron spectroscopy from QCD, the underlying theory.
Quarks
were first postulated as constituents of hadrons to describe
qualitatively
the spectroscopy of mesons and baryons containing the three lightest
$u$,
$d$ and $s$ quark flavours.\refmark{\NQM}
Subsequently, it was realized that the
short-distance properties of strongly-interacting matter could be
described
exactly in terms of current quarks and the asymptotic freedom of QCD.
The
phenomenological successes of current algebra\refmark{\CA}
and chiral symmetry\refmark{\GOR}
implied
that the current light ($u$, $d$, $s$) quarks must be much lighter than
the original constituent quarks, and the relation between current and
constituent light quarks awaits clarification.
\refmark{\GM-\EFK}
The distinction between
current and constituent heavy quarks $(c,b,t)$ is also not clear, but is
not
so crucial. QCD-inspired potential models work well for mesons made
out of
heavy quarks, and lattice techniques provide fair understanding of the
effective heavy quark--anti-quark potential.

The mystery of the relationship between light current and apparently
the heavier constituent quarks is only deepened by the successes of
calculations
of baryon properties made using the Skyrme
model\refmark{\Skyrme,\SM},
a soliton in the low-energy
chiral approximation\refmark{\WittenLewes}
 to QCD in terms of bosonic matrix variables.
Constituent
quarks do not appear in the Skyrme model, their r\^oles being usurped by
coherent states of current quarks.

In $1 + 1$ dimensions, the spectrum and interactions of mesons
in QCD$_2$ were first discussed in the framework of the
large-$N_c$ expansion.\refmark{\tH-\BESW}
For baryons, non-Abelian bosonization methods\refmark{\WittenBoson}
applied to QCD$_2$\refmark{\Gonzales} have made it possible to
obtain the low-lying spectrum\refmark{\DFS-\FK}
in the case of an unbroken light
flavour
symmetry, again without any reference to the idea of constituent quarks.
 \foot{The one-flavour case in the discretized light-cone
formalism is discussed in ref.~[\HBP].}
More
recently, explicit asymptotic static soliton solutions of the bosonized
heavy
quark theory have been exhibited.\refmark{\EFK}
These have the quantum numbers of
quarks and an infinite energy associated with a colour flux tube of
infinite length.
\foot{There are also qualitative and group-theoretical
indications that such a mechanism
could be responsible for appearance of constituent quarks
in QCD$_4$, but the relevant dynamics is as yet
unknown.\refmark{\KaplanI}}

In this paper we extend this approach by exhibiting static
soliton solutions of QCD$_2$ that have the quantum numbers of  baryons
and mesons.
These new solutions are colour singlets and have finite energy.
The solutions with baryon number zero are bound states
of the quark and anti-quark solitons, while  those with non-zero
baryon number are bound states of $N_c$
quark solitons, corresponding to mesons and baryons, respectively.
They provide a
theoretical laboratory in which the concept of a constituent quark can
be
dissected. They also provide insight into the QCD description of
heavy-light $Q\bar q$ mesons such as the $D$ and $B$,
and baryons with one or two heavy quarks,
to which the
previous heavy-quark effective potential and light-quark chiral
approaches have not been applicable. We show that the $D$ and $B$ mesons
are likely to contain OZI-evading densities of quark--anti-quark pairs
that
are absent in the na\"\i ve constituent quark description, and could
play
observable r\^oles in their dynamics and decays.

The paper is organized as follows.
In section 2 we show how mesons emerge as electron-positron solitons in
QED$_2$, developing intuition for the QCD$_2$ case discussed in
section 3. Section 4 discusses explicit meson and baryon solutions
in QCD$_2$,
in terms of quark and anti-quark solitons. Such solutions turn out to
exist only when the  fundamental fields of
the  bosonized Lagrangian  belong  to $U(N_c{\times}N_f)$
rather  than to \hbox{$SU(N_c) \times U(N_f)$.}
Section 5 discusses their semi-classical quantization.
Section 6 contains comments on $D$- and $B$-meson physics,
and section 7 is a summary and outlook.
Formal aspects of mass splitting and normal ordering ambiguities
are discussed in an Appendix.

\chapter{Mesons from solitons in QED$_2$}

Some of the interesting nonperturbative
phenomena in QCD$_2$ have close analogues
in QED$_2$ and are easy to derive, once the bosonized form of
the Lagrangian is known. In this section we present a rederivation
of the relevant results obtained long
ago, via Abelian bosonization, by Coleman\refmark{\Coleman},
and add some new results of
our own, namely explicit solutions
in the case of broken flavour symmetry. We believe the reader will
find this section useful for developing physical intuition
for the discussion of QCD$_2$ to follow in the next section.

The Lagrangian of multi-flavour massive QED in two dimensions is
$$ \eqalign{
\LL = \sum_k \psibar_k (i &\Dslash - m_k) \psi_k
-{1\over4} F_{\mu\nu} F^{\mu\nu} \crr
i D_\mu &= i \partial_\mu - e A_\mu
}\eqn\QEDI$$
where $k$ is the flavour index, $A_\mu$ is the gauge
potential, and
$$F_{\mu\nu} = \partial_\mu A_\nu - \partial_\nu A_\mu
\equiv \epsilon_{\mu\nu} F\eqn\QEDIa$$
The bosonized version of \QEDI\ reads\refmark{\Coleman}
$$ \eqalign{
\LL = {1\over2} \sum_k (\partial_\mu \chi_k)^2
+{1\over2} F^2 &+ {e\over \sqrt{\pi}} F \sum_k \chi_k
+\sum_k {C\over\pi} m_k \mu_k N_{\mu_k} \cos \sqrt{4\pi} \chi_k \crr
\psibar_k \gamma_\mu \psi_k & = - {1\over\sqrt{\pi}}
\epsilon_{\mu\nu} \partial^\nu \chi_k\crr
\psibar_k \psi_k & = - {C\over\pi} \mu_k N_{\mu_k}
\cos \sqrt{4\pi} \chi_k\crr
Q=\int_{-\infty}^{\infty} d x \sum_k
\psi^\dagger_k \psi_k & = {1\over\sqrt{\pi}}
\sum_k [\chi_k(\infty) - \chi_k(-\infty)]
}\eqn\QEDII$$
where $Q$ is the total electric charge,
$C = 1/2~ e^\gamma \simeq$ 0.891 ($\gamma$ is Euler's constant),
$N_{\mu_k}$ stands for normal ordering
\foot{For a detailed discussion of normal ordering in the presence
of different masses, see the Appendix.}
with respect to a scale $\mu_k$,
and we take the $\theta$
vacuum angle\refmark{\Coleman} to be zero, for simplicity.

We integrate out $F$ and take $\mu_k = m_k$,
and then ``absorb"
a factor of $C/\pi$ in a redefinition of $m_k^2$.
Then
the Lagrangian \QEDII\ becomes
$$ \LL = {1\over2} \sum_k (\partial_\mu \chi_k)^2
- {e^2\over 2{\pi}} (\sum_k \chi_k)^2
+\sum_k m_k^2 \cos \sqrt{4\pi} \chi_k
\eqn\QEDIII$$
The equations of motion in the static case read
$$ \chi_k^{\prime\prime} - 4 \alpha (\sum_l  \chi_l) -
\sqrt{4\pi} m_k^2 \sin \sqrt{4\pi}\chi_k = 0 \eqn\QEDIV$$
where $\ds\alpha \equiv {\ds e^2\over 4\pi \ds}$. The static potential
is given by
$$V = 2 \alpha (\sum_l  \chi_l)^2 +
\sum_k m_k^2 (1- \cos \sqrt{4\pi} \chi_k) \eqn\QEDV$$
Multiplying \QEDIV\ by $\chi_k^{\prime}$, summing over $k$ and
integrating, we get
$${1\over2}\sum_k{\chi_k^{\prime}}^2-V(x) = \hbox{(const.)}\eqn\QEDVI$$
We are looking for finite energy static solutions, therefore
the gradient terms must vanish asymptotically,
$$\chi_k^{\prime}(\pm\infty)=0 \eqn\QEDVII$$
so that
$$V(\infty) = V(-\infty)\eqn\QEDVIII$$
Taking $\chi_k(-\infty)=0$,  we see that $V(-\infty)=0$,
and hence $V(\infty)=0$. Thus
$$ \eqalign{
\sum_l \chi_l(\infty) & = 0 \crr
\cos \sqrt{4\pi} \chi_k(\infty) &= 1 }\eqn\QEDIX$$
We thus see that only states with zero total charge $Q$
are allowed, which is what one expects, since QED$_2$ is
confining.
 From \QEDIX\ it follows
$$ \chi_k(\infty) = \sqrt{\pi}\, n_k, \,\qquad n_k=0,\pm1,\pm2
,\dots
\eqn\QEDX$$
For two flavours, from \QEDIX\ and \QEDX, taking the lowest
non-trivial $n_k$'s, namely $n_1=1,n_2=-1$, we obtain
$$ \eqalign{
   \chi_1(\infty) = \phantom{-} \sqrt{\pi} \crr
   \chi_2(\infty) =          -  \sqrt{\pi}
}\eqn\QEDXI$$
The boundary conditions \QEDXI\ correspond to a meson built
out of a soliton and an anti-soliton.
Equations \QEDIV\ with boundary conditions \QEDXI\
can be solved explicitly.
When $m_1 = m_2$ it is easy to see that
$\chi_1(x)=-\chi_2(x)$ and the ``string tension"
term proportional to $\alpha$
in \QEDIV\ vanishes, leading to two ``mirror" decoupled
sine-Gordon equations for $\chi_1$, $\chi_2$. When $m_1\neq m_2$,
a solution can be found numerically. It is particularly interesting
to examine the solutions for widely unequal masses. Some representative
solutions are shown in Fig.~1. It is worthwhile pointing out that
for $\alpha \ll m_1^2, m_2^2$, the widths of $\chi_1$ and $\chi_2$
are governed by $1/m_1$ and $1/m_2$, respectively. The lighter
``quark", say $\chi_2$,
will be more influenced by the string tension, getting
heavier with increasing $\alpha/m_2^2 $.
 This is just the intuitive picture
we usually associate with a light current quark getting a constituent
mass of the order of the gauge scale.

Let us now consider the case $m_1\rightarrow\infty$. Then,
in \QEDIV\ with $k=1$, we can neglect the $\alpha $ term and
get a free soliton of mass $m_1$, which in the limiting case
tends to a theta function
$$ \chi_1(x) \Botrel{\longrightarrow}\over{m_1\rightarrow\infty}
\sqrt{\pi}\,  \theta(x)\eqn\QEDXII$$
as follows from
the explicit form of the sine-Gordon solution,
$$\chi_1={2\over\sqrt{\pi}} \tan^{-1}\exp\left(\sqrt{4\pi} m_1 x\right)
\eqn\QEDXIIa$$
Then, for the light flavour, $k=2$,
$$\chi_2^{\prime\prime} -4 \alpha \chi_2 - \sqrt{4\pi} m_2^2
\sin \sqrt{4\pi} \chi_1 =  4\alpha \sqrt{\pi}\theta(x)
\ .\eqn\QEDXIII$$
We see that the light ``anti-quark" field $\chi_2$ feels a point-like
``source" term due to the heavy ``quark" $\chi_1$.
In the following sections we shall demonstrate that
a very similar phenomenon occurs in QCD$_2$.

\chapter{Hadronic Solitons}

Two non-Abelian bosonizations of QCD$_2$ have been developed,
one in terms of
$SU(N_c)\times U(N_f)$ bosonic variables\refmark{\DFS}
where $N_c$ is the number of
colours and $N_f$ is the number of flavours,
and the other\refmark{\FS} in terms of
$U(N_c{\times}N_f)$ bosonic variables.
\foot{The specific case
of $SU(N_c)$, $N_f=2$ has also been considered in a mixed
Abelian -- non-Abelian formalism in Ref.~\Gepner.}
We will now show that
in the  $U(N_c{\times}N_f)$ scheme, but not in the
$SU(N_c)\times U(N_f)$ scheme,  there are
static  solutions
that have  the quantum numbers of  baryons and mesons,
constructed out  of quark and anti-quark  solitons.  In isolation
the latter  solitons have  infinite energy, corresponding  to the
presence of a string carrying  the non-singlet colour flux off to
spatial infinity.
When  $N_c$  solitons  of  this  type  are  combined,  a  static,
finite-energy, colour  singlet solution is  formed, corresponding
to a baryon. Similarly, static  meson solutions are formed out of
a soliton and  an anti-soliton of different  flavours.  The stability
of the mesons against annihilation is ensured by  flavour conservation.

The effective exact  $U(N_c{\times}N_f)$ flavour symmetric
bosonic action with gauge fields integrated out reads\refmark{\FS}
$$\SeffU = S_0[u] + {e_c^2 N_f \over 8 \pi^2}
\int d\,^2 x \Tr \lsquare \dminus^{-1} \left( u \,\dminus
u^\dagger\right)_c \right]^2
+ \mprime^2 N_\mtilde \int d\,^2 x \Tr \left( u + u^\dagger\right)
\eqn\I
$$
where\refmark{\WittenBoson}
$$
S_0[u] =
{1\over8\pi} \int d\,^2 x \Tr (\partial_\mu u)
(\partial^\mu u^\dagger)
+{1\over12\pi} \int_B d\,^3 x \Tr\epsilon^{ijk}
(u^\dagger \partial_i u)
(u^\dagger \partial_j u)
(u^\dagger \partial_k u)
\eqn\II
$$
$e_c$ is the strong coupling constant (having dimensions
of mass in 1+1 space time),
$u$ is an element in $U(N_c{\times}N_f)$,
the subscript $c$ denotes projection onto the colour
part, i.e. averaging
over flavour and subtracting the $U(1)$ part, and
$$\mprime^2 = m_q \mtilde\,{C\over2\pi}\,.
\eqn\III
$$
where $\tilde m$ is a normal ordering scale,
and $N_{\tilde m}$ stands for normal ordering with
respect to $\tilde m$ (see the discussion in the Appendix).
We shall look for solutions with $u$ in diagonal
form (see the discussion in ref.~[\DFS]),
$$\eqalign{
u_{\alpha\alphaprime j\jprime} & =
\delta_{\alpha\alphaprime}\,\delta_{j\jprime}
e^{\displaystyle - i\spi\chiaj} \crr \alpha&=1,\dots,N_c\cr
j&=1,\dots,N_f\cr}
\eqn\IV
$$
The Wess-Zumino term vanishes for either static or diagonal solutions.
In general,
for non-equal masses, in the static case, %
$$\eqalign{
\SeffU =
& -{1\over 2} \sum_{\aj} \int \left\(\chipaj(x) \right\)^2
- 2\ac \sum_\alpha \int \left\(\sum_l \chial - {1\over N_c } \sum_\bl
\chibl \right\)^2 \crr
& +  \sum_\aj \int m_j^2 \cos \spi \chiaj}
\eqn\V
$$
where $\ac=e_c^2/4\pi$ and
$m_j$ stands for the $j$-th mass.
The relation of $m_j$'s to the ``current" quark masses
and to $\alpha_c$, as obtained through
the normal ordering, is discussed in the Appendix.

The Hamiltonian density is
$$ \eqalign{
H &= {1\over2} \sum_{\aj} \left( \chipaj\right) ^2 + V, \crr
V &= 2\ac \sum_\alpha \left\(\sum_l \chial - {1\over N_c } \sum_\bl
\chibl \right\)^2
+  \sum_\aj m_j^2 \left\(1 - \cos \spi \chiaj \right\)}
\eqn\VI
$$
where we have added a constant to $V$, to make $V$ vanish for
\ $\chiaj=0$.
The equations of motion are
$$ \chi^{\prime\prime}_{\aj}
- 4 \ac \left\( \sum_l \chial - {1\over N_c}
\sum_{\bl} \chibl \right\) - \sqrt{4\pi} m_j^2 \sin \spi \chiaj = 0
\eqn\VII
$$
To obtain an integral of motion, we multiply by $\chi '_{\alpha_j}$ and
sum
over $\alpha , j$, to obtain in the static case
$$ \sum_\aj - {1\over2} \left(
\chipaj\right) ^2 + 2\ac \sum_\alpha \left\(\sum_l \chial - {1\over N_c }
\sum_\bl
\chibl \right\)^2
+ \sum_\aj m_j^2 \left\(1 - \cos \spi \chiaj \right\)
= \hbox{(const.)}
\eqn\VIII
$$
Namely, the left-hand side of \VIII\ is independent of $x$.
For  static solutions, the existence of such an
$x$-independent integral of motion is the
analogue of energy conservation for solutions evolving in time.

If we choose the boundary condition $\chi
_{\alpha j}
(-\infty) = 0$, then the constant on the right-hand side of \VIII\
vanishes, and
hence also for $\chi_{\alpha j}(+\infty )$ (denoted hereafter simply by
$\chi_{\alpha j}$), we must have
$$
2\ac \sum_\alpha \left\(\sum_l \chial
- {1\over N_c } \sum_\bl \chibl \right\)^2
+  \sum_\aj m_j^2 \left\(1 - \cos \spi \chiaj \right\)
\Bigg\vert_{+\infty} = 0
\eqn\IX
$$
Note that this condition is also obtained by requiring
finite energy, as follows from eq.~\VI.
We infer from \IX\ that
$$\eqalign{
{1\over\sspi} \chiaj &= \naj \qquad\hbox{integers} \crr
\sum_l \chial & = \sspi \sum_l \nal = \sspi n \qquad
\hbox{independent of $\alpha$}}
\eqn\X
$$
The baryon number of any given flavour $ l$ is given by
$$ B_l = \sum_\alpha \nal
\eqn\XI
$$
Combining eqs.~\X\ and \XI\ we get the total baryon number
$$
B = \sum_l B_l = nN_c
\eqn\XIII
$$
which clearly is an integer multiple of $N_c$.

\chapter{Explicit solutions and bosonization schemes}

For the mesonic solutions, we need $B = 0$ and hence $n = 0$.  Let us
consider
the case of a meson containing a quark of flavour $l=1$ and an
anti-quark
of flavour $l=2$, and no other constituents. Thus, we need
$$\eqalign{
 \sum_l      \nal &= 0 \crr
 \sum_\alpha \nal &=
 \left\{\matrix {\pp 1 & l=1 \cr
                    -1 & l=2 \cr
                 \pp 0 & l \ge 3 \cr} \right.}
\eqn\XIV
$$
Note that in the product scheme, i.e. with $h$ in $SU(N_c)$
and $g$ in $U(N_f)$, we would have
$$\eqalign{\nal & = p_\alpha + q_l \crr
\sum_\alpha p_\alpha & = 0
}\eqn\ProductScheme$$
Then
$$ N_f p_\alpha + \sum_l q_l = 0 \eqn\PSII$$
which implies that $p_\alpha$ is independent of $\alpha $ and
hence must be zero. Thus
$$\nal = q_l \qquad \hbox{integers}\eqn\PSIII$$
$$\eqalign{
 \sum_l  q_l &= 0 \crr
N_c \,q_l &=
 \left\{\matrix {\pp 1 & l=1     \cr
                    -1 & l=2     \cr
                 \pp 0 & l \ge 3 \cr} \right.
}\eqn\PSIV$$
which is impossible.
Thus there are no solutions with mesons as quark--anti-quark
states in the ``product scheme" of $SU(N_c)\times U(N_f)$.

Let us now give an example of a solution in the
$U(N_c{\times}N_f)$ scheme. In view of the discussion above,
such a solution will obviously have components in what we
called $l$ in ref.~[\FS], namely a non-factorizable part in
the colour-flavour space.

As obtained above, the asymptotic boundary conditions are:
$$\eqalign{
\chi_{\alpha j}(-\infty)&=0\crr
\chi_{\alpha j}(+\infty)&=\sqrt{\pi} n_{\alpha j}
}\eqn\LeftRightBC$$
where the set $\{n_{\alpha j} \}$ must satisfy the constraints
\XIV.
A possible solution is
$$ \eqalign{
n_{11} &=\pp 1 \crr
n_{12} &=-1
}\eqn\XV
$$
with all other $\naj$ being zero.
Having specified the asymptotic boundary conditions
at $x \rightarrow \pm \infty$, we must now
see whether a solution exists for all $x$.
In general such solutions can only be found numerically.\foot{We have
used the subroutine package COLSYS.\refmark{\COLSYS}}
The case of an exact $SU(N_f)$ symmetry, i.e.
equal quark masses, is an exception where an explicit
analytical solution can be found:
$$\chi_{11}(x)=-\chi_{12}(x)={2\over\sqrt{\pi}} \,\tan^{-1}
\left[\exp\left(\sqrt{4\pi}m\,x\right)\right],\eqn\ExactSoln$$
with all others vanishing identically. In order to show that
this is a solution, we note that with \ExactSoln\
the string tension term proportional to $\alpha_c $
in \VII\ vanishes identically, and the
equations for the various $\chi_{\alpha j}$ decouple, yielding
a set of independent sine-Gordon equations with equal masses.
It is then obvious that with the boundary conditions
\LeftRightBC, \XV,
$\chi_{11}(x)$ is the usual sine-Gordon soliton.

Recently there has been much discussion in the literature of the
so-called heavy quark symmetry\refmark{\WI-\GrinsteinRev}.
Here this symmetry manifests
itself in a rather clear way.
Consider a $Q\bar{q}$ meson made out of a heavy quark $Q$ and a
light anti-quark $\bar{q}$, such as the $D$- or $B$-mesons.
When $m_Q$ is much larger than the
scale of the theory, $m_Q \gg e_c$, its profile tends to a theta
function,
$$ \chi_{1 Q}
\Botrel{\longrightarrow} \over{m_Q/e_c \,\rightarrow\, \infty}
\sqrt{\pi}\,\theta(x)
\eqn\HeavyProfile$$
as in the QED$_2$ case,
while the baryonic current of the heavy quark
tends to a delta function,
$$ J^B_{Q}={1\over\sqrt{\pi}}\,\sum_{\alpha} \partial_x\chi_{\alpha Q}
\Botrel{\longrightarrow} \over{m_Q/e_c \,\rightarrow\, \infty}
\delta(x)
\eqn\HeavyCurrent$$
Thus the heavy quark acts as a static colour source,
while  the profile $\chi_{\alpha q}$ of the light
quark becomes independent of $m_Q$, as shown in Fig.~2.
The physics of $B$ and $D$ mesons in the context of
the heavy quark symmetry will be discussed in more detail
in section~6.

The presence of the static colour source
\HeavyCurrent\ makes the total
energy of the $Q\bar{q}$ system finite.
 From \VI, \XI, \LeftRightBC\ and \XV\ we see that
had there been no other colour source, the
light flavour profile $\chi_{1 \bar{q}}$  on its own would
correspond to a configuration with  a net baryon number ${-}1$,
one ``unit" of colour charge and
a finite energy density per unit length, associated with
a colour flux tube of infinite length,
$$ \eqalign{\chi_{1 \bar{q}}
&\Botrel{\longrightarrow} \over{x\,\rightarrow\, \infty}
- \sqrt{\pi}\crr
V_{\bar{q}}(x)
&\Botrel{\longrightarrow} \over{x\,\rightarrow\, \infty}
2 \pi \alpha_c \left( 1 - {1\over N_c} \right)^2
}\eqn\LightProfile$$
resulting in the total energy being infinite.
This is reminiscent of
the single quark solution discussed in ref.~[\EFK].
It gives precise meaning to the intuitive concept
of quark confinement: isolated quarks have infinite energy
because flux conservation forces them to emit a flux tube
which has no sink to absorb it.

Multi-quark baryonic solutions can be obtained in a similar way
to the meson solutions.
For example, taking $N_c=3$ and $B=3$
(a 3-quark state\foot{In our normalization
a single quark carries one unit of baryon number.})
we find $n=1$ ({\sl cf.} eq.~\XIII). One possible solution is
$$ n_{11} = n_{21} = n_{31} = 1 \eqn\FrozenColour$$
Corresponding to a ``$uuu$"-like baryon, i.e. a ``$\Delta^{++}$".
When quark masses are equal, $m_1=m_2=\dots=m_{N_f}$,
this solution coincides with one  found earlier in the ``product
 scheme"\refmark{\DFS},
in which the colour part is ``frozen", i.e. the string tension
vanishes identically. The vanishing of string tension in \FrozenColour\
is caused by a mechanism  similar to the one occurring in the meson with
equal quark masses described  above,  leading to
$$\chi_{11}(x)=\chi_{21}(x)=\chi_{31}(x)={2\over\sqrt{\pi}} \,\tan^{-1}
\left[\exp\left(\sqrt{4\pi}m\,x\right)\right],\eqn\ExactBaryon$$
This baryon solution is plotted
in Fig.~3(a).

There is an important difference, however, between the meson and
the baryon cases. The vanishing of the string tension term in the
meson is a manifestation of the fact that the chromoelectric fluxes
of the quark and anti-quark
cancel each other. This phenomenon has its
counterpart in QED$_2$\refmark{\Coleman}, as discussed in section 2.
The cancellation
of fluxes of $N_c$ quarks has no such counterpart and can only occur
in a non-Abelian theory. Another difference between the baryonic and
mesonic solitons is that the latter do not exist in the ``product
scheme".

When the boundary conditions \FrozenColour\ are taken together with
non-equal quark masses, we still obtain a ``$uuu$"-like baryon,
with the classical solution \ExactBaryon, as
in the ``product scheme". The quantum fluctuations, however,
must be treated differently, due to flavour symmetry breaking
(see the Appendix for additional discussion).

An intrinsically new 3-quark solution occurs when
more than one flavour appears in the nontrivial solution, for
example,
$$ n_{11} = n_{22} = n_{33} = 1 \eqn\genuineSolution $$
This corresponds to  a baryon in which each quark has a different
flavour.
Such a solution is particularly interesting when quark masses are taken
to be non-equal,  corresponding to a ``uds"-like baryon, or to
a baryon in which one quark is much heavier than the QCD scale,
such as the  $\Sigma_c$ or $\Sigma_b$,
\foot{In 1+1 dimensions
there is no spin and therefore baryons must be
in the completely symmetric representation of $SU(N_f)$,
as the space part of the wave function is symmetric for
lowest energy states.
Thus with $N_c=3$ and three light flavours
there is only  decuplet and no octet,
and there is no analogue of $\Lambda_c$.}
again serving as a theoretical laboratory for the study of
the heavy quark symmetry discussed earlier.
\foot{For a different approach to heavy-quark baryons in the
chiral soliton framework, see ref.~\Rho.}

One can also
study baryons  containing two heavy  quarks, as shown  in Fig.~3(b).
For $N_c=3$, the light quark distribution in such a baryon is the
same as in a $\bar{Q} q$  meson.  The physical reason for this is
that the  two heavy quarks  are essentially at  rest and act  as a
static colour source.  They are  colour triplets and combine to a
colour  anti-triplet,
${\bf  3} \otimes  {\bf 3}  \rightarrow {\bf 3}^{*}$,
i.e. the effective field seen by the light quark $q$ in a $QQq$
baryon is that of a static anti-quark.
This gives precise meaning to the concept of constituent quark
in QCD$_2$. The $QQq$ case is particularly clear, since
this type of baryon contains only one light constituent quark, while
in $Qqq$ or $qqq$ baryons there are two or three such objects,
 superimposed nonlinearly.
A solution of the $QQq$
type with $m_1 \ll m_2 \ll m_3$ is shown in Fig.~3(c).

\chapter{Semi-classical quantization}

As has already been mentioned, one of the specific
applications
which motivated this study was that to mesons containing just one heavy
quark
$Q$. In the constituent quark language, these would be described as
$Q\bar q$
mesons, where $\bar q$ represents some light constituent quark $(u,d$ or
$s$).
Examples include the $D,D_s$ and $B$ mesons. To describe such mesons
within our
QCD$_2$ approach, we need to consider a quark mass matrix with one heavy
eigenvalue $M \gg e_c$,
and $N_f-1$ (typically three) light eigenvalues $m \ll e_c$.
In such a case, the light quark degrees of freedom should be quantized
semi-classically, as was already done for baryons made out
of light current quarks.\refmark{\DFS}
The resulting lump is the best QCD model we can derive
for the concept of a light constituent quark. However, clearly it is a
coherent
state containing in some sense an infinite number of current quarks,
at least in the massless limit.

When discussing quantum fluctuations around the classical solution
$u_c(x)$, we write
$$ U=A u_c(x)A^{-1} \eqn\FLUCTa $$
with $A$ unitary. Now, we take $A$ to be a product
$$ A=A_c A_f\eqn\FLUCTb $$
where $A_c$ acts only in the colour space and $A_f$ in flavour space.
This  factorization is  implied  by the  fact  that although  the
bosonic variables $u(x)$ belong  to $U(N_c{\times}N_f)$, the full
theory  is {\sl  not}  $U(N_c{\times}N_f)$  invariant.  Only  the
free-fermion theory has the full $U(N_c{\times}N_f)$ symmetry, and
gauging of the colour part  results in this symmetry being broken
down to $SU(N_c){\times}U(N_f){\times}U(1)$.
If we take $A_c$ to be a function of $x_-$ only,
we see that it will not change the interaction term in
$\SeffU$ of eq.~\I, thus having no $\alpha_c$ effects.
Thus we use ``light-cone quantization" for the colour part.
For $A_f$ we consider a
function of $t$ only, thus reducing the treatment of
quantum fluctuations to that
which has already been treated in ref.~[\DFS],
for the equal mass case.  The unequal mass
case is discussed in the Appendix.

Taking  the colour part $A_c$ to be a function of
$x_-\ $, while the flavour  part $A_f$ is taken to be a function
of $t$, might turn out  to be  problematic.
A  more  natural choice would have been
to take both $A_c$ and $A_f$ as functions of $t$, since we are
performing the quantization around a static classical
solution, $u_c=u_c(x)$.

In any case, we expect
that for equal quark masses  the fluctuations in the colour space
will not  influence ratios  such as  flavour content  of hadrons
(sec.~6),  since the  physical states  are colour  singlets.  The
situation  might  change,  though,  when  masses  are  different.
The flavour content will  then in general depend on  the mass ratios,
and the different masses could in general be influenced
differently by
the colour fluctuations.  This is  because the flavours which are
light compared with the typical scale of the  theory are in
the strong-coupling  regime, and  will be  more affected  than the
flavours which are much heavier than the dynamical
scale of the theory. In our case, we will be interested in the
$u$, $d$ and $s$ contents in the approximation where they are
all very light.

\chapter{Comments on $D$ and $B$ physics}

In the previous sections we have described QCD$_2$ solitons which
could serve  as models for  $Q\bar q$ mesons  such as the  $D$ or
$B$.  In  addition to giving some  insight into the concept  of a
constituent quark from the point of QCD, this study may also give
some new  insights into the dynamics  and weak decays of  $D$ and
$B$  mesons.
\foot{$Q\bar{q}$ mesons in QCD$_2$ have also been
recently studied in the large-$N_c$ limit.\refmark{\GrinsteinMende}}
 In  particular, we  would  like to  comment on  the
existence  and possible  phenomenological  r\^ole of  non-valence
quarks in the $D$ and $B$ meson wave functions.

There are various phenomenological indications that the proton wave
function
contains a significant density of non-valence $\bar ss$ quarks.
Moreover, their
abundance relative to $\bar{u}u$ and $\bar{d}d$ quarks is qualitatively
reproduced by Skyrme model soliton calculations. The relative abundances
of
$\bar ss, \bar uu$ and $\bar dd$ have also been calculated in baryonic
QCD$_2$ solitons,\refmark{\FK}
and found to be qualitatively similar to the
QCD$_4$ results.\refmark{\DonoNappi-\Weigel}
In this
chapter we invert the logic: QCD$_2$ mesonic solitons contain calculable
non-valence quark densities, and we would expect QCD$_4$ mesonic
solitons and
hence physical $D$ and $B$ mesons to contain similar non-valence quark
densities.

The calculation of the different light quark densities in a QCD$_2$
mesonic
soliton parallels very closely that in a baryonic soliton:\refmark{\FK}
$$
{{\langle M\vert\bar q_iq_i\vert M\rangle}\over{\langle  M\vert\bar
q_jq_j\vert
M\rangle}} = {{\ll z^*_iz_i \gg}\over {\ll z^*_jz_j \gg}}
\eqn\un
$$
where $\ll\quad\gg$ denotes an average over the collective coordinate
representation of the mesonic soliton. In the physical case with one
heavy and
three light quarks we find for a $Q\bar q_1$ meson:
$$
\langle M\vert\bar q_1q_1\vert M\rangle : \langle M\vert\bar q_2q_2,\bar
q_3q_3\vert M \rangle = 2:1
\eqn\two
$$
This parallels the previous result \refmark{\FK}
for the sea- and valence-quark content of baryons in the semi-classical
quantization of QCD$_2$ with $N_l$  light flavours and $N_c$ colours.
For a baryon $B$ containing $k$ light valence quarks
of flavour $v$
$$\VEV{\bar{v} v}_B={k+1\over N_l + N_c} \eqn\GenCaseV$$
and
$$\VEV{(\bar{q}q)_{sea}}_B={1\over N_l+N_c},\eqn\GenCaseSea$$
where
$(\bar{q}q)_{sea}$ refers to the non-valence quarks in the
baryon $B$, and the total flavour content is normalized to 1.
Equations \GenCaseV\ and \GenCaseSea\ suggest
an ``equipartition" for valence and sea,
each with a content of \hbox{$1/(N_l+N_c)$.}
Since the semi-classical quantization of light flavours
in $\bar{Q}q$ mesons parallels that in the baryons,
the ``equipartition"
result applies to the light
flavour content of $\bar{Q}q$ mesons as well.
This implies, for example, that
$$
\langle D\vert\bar ss\vert D\rangle / \langle D\vert\bar uu\vert
D\rangle =
{{1}\over {2}}
\eqn\three
$$
and similarly for $B$ mesons. We might expect
the ratio \three\ to be
qualitatively similar, though possibly smaller by about a factor 2,
\refmark{\FK}
 for the realistic case of QCD$_4$ with light
flavour
$SU(3)$ breaking.\refmark{\DonoNappi-\Weigel}

The presence of significant amounts of $\bar ss$ quarks in the $D$ and
$B$
mesons implies that the na\"\i ve OZI rule forbidding disconnected quark
diagrams can be evaded.\refmark{\EGK}
This might also have implications for $D$ and
$B$
production dynamics, but here we only emphasize some possible
implications for
$D$ and $B$ decays.

\pointbegin
Annihilation diagrams: $c\bar s\rightarrow u\bar d$ could be more
important for
$D^0$ and $D^+$ mesons than is normally supposed when only the $D_s$
wave
function is assumed to contain strange quarks.
\point
The final states from $D$ and $B$ decays could contain more strange
particles
than is normally supposed when $\bar ss$ pairs need to pop out of the
QCD
vacuum or be created at the weak vertex. This could help explain the
surprisingly large \refmark{\DOZItheory,\DOZIdata,\Anjos}
branching ratios for \hbox{$D^0\rightarrow\phi K^0$} and
{$D^+\rightarrow \phi K^+$.} This observation could also have implications
for
attempts to estimate the ratio$\vert V_{cs}/V_{cd}\vert$ of
Kobayashi-Maskawa
matrix elements on the basis of strange final states.
\refmark{\Adamovich} It might also have
implications for the ratios of $D\rightarrow K\bar K$ and $\pi\pi$ final
states.
\point
The  $\bar ss$ pairs could provide an additional source for
$B\rightarrow \phi + X$. At this time it is premature to compare
this with the data.

Detailed investigations of these possibilities should await more
realistic
calculations in QCD$_4$, however.

\chapter{Summary and outlook}

We have shown in this paper that the spectrum of QCD$_2$ includes
finite-energy
mesonic and baryonic solitons which contain one heavy quark.
These solitons can be regarded
as bound states of the infinite-energy single-quark solitons that we
found
previously. They provide meaning for the previously fuzzy concept of a
constituent quark, at least in QCD$_2$.
A particularly interesting application is to the study of mesons
and baryons containing both heavy and light quarks.
A constituent light quark is
seen to be
a semi-classical coherent state containing an indefinite number of light
$\bar
qq$ pairs, among which non-valence flavours have as much as one half of
the
abundance of the valence flavour. This observation could have
phenomenological
implications for the dynamics and weak decays of $D$ and $B$ mesons.

The next step in the programme of developing accurate QCD descriptions
of
constituent quarks and $Q\bar q$ mesons is to extend the analysis of
this paper
to the realistic case of four dimensions. This may be possible for the
lowest-lying $Q\bar q$ mesons if they are describable by spherically
symmetric
wave functions, which could be analyzed using an effective
two-dimensional
field theory in $(r,t)$ coordinates. We are now investigating this
possibility.

\vskip1cm
\vbox{
\noindent
{\bf ACKNOWLEDGEMENTS}

One of us (J.E.) thanks the Minerva Foundation, the Weizmann Institute
and Tel
Aviv University for their kind hospitality during the early stages of
the work
described here.
The research of M.K.  was supported in part
by the Einstein Center at the Weizmann Institute and
by the Basic Research Foundation administered by the
Israel Academy of Sciences and Humanities
and by grant No.~90-00342 from the United States-Israel
Binational Science Foundation(BSF), Jerusalem, Israel.
M.K. would like to thank Shimon Yankielowicz for useful
discussions.
}

\equanumber=0
\appendix
We discuss here the more formal aspects of mass splitting.
In the following we will focus mainly on the bosonization in the
product scheme, since the technicalities are simpler there, while
the conclusions are the same as in the
$U(N_c{\times}N_f)$ scheme.
We recall the initial bosonic action
in the product scheme\refmark{\DFS}
$$\eqalign{
\tilde S\left\(g,h,A_\mu\right\)&=N_c \, S\left\( g\right\)
+ N_fS\left\(h,A_\mu\right\)
-{1\over 2e_c^2}\int d^2 x\Tr\,F_{\mu\nu}F^{\mu\nu}\crr
&+C\tilde m N_{\tilde m}
\int d^2x\left\( \Tr\left( M\,g\right) \Tr\,h+h.c.\right\)}\eqn\amiI$$
where
$g \in U(N_f)$,
$h \in SU(N_c$),
$A_\mu$ is colour gauge field,
$M=\diag(m_1,\dots,m_{N_f})$ is the
mass matrix, $m_i$ are the quark masses, $N_\mtilde$\ \ stands for
normal ordering with respect to \mtilde, and $F_{\mu\nu}$ is the
usual non-Abelian field-strength tensor.
Now define the ``dimensionless mass matrix" $D$ as
$$ D ={1\over m_0}M \eqn{\m}$$
where $m_0$ is an arbitrary mass parameter.
In the gauge $A_{-}=0$, integrating out $A_{+}$, and then
taking the strong-coupling limit, $e_c\gg$ all $m_i$, we get
\refmark{\DFS}
$$\tilde S\left\( g\right\)=N_cS\left\( g\right\)+m^2N_m
\int d^2 x \, Tr\left\( D(g+g^\dagger )\right\) \eqn{\eqsc}$$
$$m=\left\( C N_c m_0 \left(e_c^2N_f\over 2\pi\right)^{{1\over 2}p}
\right\)^{1\over p+1} \eqn\amiII$$
$$p={{N_c^2-1}\over {N_c(N_c+N_f)}}\eqn\amiIII$$
In the $U(N_c{\times}N_f)$ bosonization scheme
the mass term, denoted M.T., has the form
$$M.T.=m_0C\tilde m N_{\tilde m}
\Tr\left\(\hat D(u+u^\dagger)\right\)\eqn\amiIV$$
where $u\in U(N_c{\times}N_f)$ and $\hat D$ is the dimensionless
mass matrix
(in flavour only).
In the strong-coupling limit the mass
term \amiIV\ exactly coincides with that of \eqsc.\refmark{\FS}
The classical low-lying solutions to the action \eqsc\ are given by
the time-independent form
$$g=\diag\left[\exp\left(-i\sqrt{4\pi\over N_c}\phi_1(x)\right),\dots,
\exp\left(-i\sqrt{4\pi\over N_c}\phi_{N_f}(x)\right)\right]\eqn{\gx}$$
yielding an action density
$$\tilde S_d\left\( g\right\) =-\int d x
\sum_{i=1}^{N_f}\left\({1\over 2}\left({d\phi_i\over d x}\right)^2
-2\tilde m_i^2\left(\cos\sqrt{4\pi\over N_c}\phi_i-1\right)\right\)
\eqn{\sdg}$$
with mass parameters
$$\tilde m_i^2={m_i\over m_0}\,m^2\eqn\amiV$$
Each $\phi_i$ has a classical solution
$$\phi_i(x)=\sqrt{4N_c\over\pi}\arctan\left\(\exp\left(\sqrt
 {8\pi\over N_c}\tilde m_ix\right)\right\)\eqn{\fix}$$
with an energy
$$E_i=4\tilde m_i\sqrt {2N_c\over\pi} \eqn{\ei}$$
The minimum classical energy solution is given by
a solution with exactly one non-trivial entry,
$$g_0(x)=\diag\left[1,1,\dots,
\exp\left(-i\sqrt{4\pi\over N_c}\phi_{N_f}(x)\right)\right]\eqn{\g0}$$
with $m_{N_f}$ chosen to be the smallest mass.

In order to quantize the system we
introduce the collective coordinates $A(t)$,
$$g(x,t)=A(t)\,g_0(x)\,A^\dagger (t),\qquad A(t)\in U(N_f)\eqn{\At}$$
Writing
$A=\left(A_1,\dots,A_{N_f-1},z\right)$
where $A_1,\dots,A_{N_f-1},z$
are column vectors, we get
$$\eqalign{
\tilde S\left\( g\right\) -\tilde S\left\( g_0 \right\) =
{1\over 2M_{N_f}}\int d t(D z)^\dagger_i(D z)_i
 +iN_c\int d t\dot z^\dagger_i&
z_i\cr -{2\pi\over M_{N_f}N_c}\int d t
\sum_{i=1}^{N_f}(\tilde m_i^2-\tilde m_{N_f}^2)\vert z_i\vert^2}
\eqn{\dsg}$$
$$D z=\dot z-z(z^\dagger\dot z)$$
$${1\over 2M_i}={N_c\over 2\pi}\int_{-\infty}^\infty
\left(1-\cos\sqrt{4\pi\over N_c}\phi_i\right)d x=
{\sqrt 2\over \tilde m_i} \left({N_c\over\pi}\right)^{3\over 2},
\qquad i=1 \dots N_f.$$
The resulting Hamiltonian is
$$H=4\tilde m_{N_f}\left({2N_c\over\pi}\right)^{1\over 2}\left\{1+\left(
{\pi\over 2N_c}\right)^2
 \left\(C_2(R)-{N_c^2\over 2N_f} (N_f-1)\right\)+\sum_{i=1}^{N_f}
 {m_i-m_{N_f}\over m_{N_f}}\vert z_i\vert^2\right\}
\eqn{\H}$$
The Hamiltonian depends on $m_0$ only through $\tilde m_{N_f}$, therefore
the overall mass scale is undetermined, and only
mass ratios are meaningful.
The quantity $\vert z_i\vert ^2$ is
proportional to the $\bar q_i q_i$ operator.
Therefore the last term in \H,
due to \hbox{$m_i \neq M_{N_f}$},
is proportional to the weighted average over flavours of the $\bar{q}q$
operators, the weights being proportional to the mass differences.
Since this term comes from quantum fluctuations around the
classical solution,
consistency with the semi-classical approximation
requires that it be very small compared to one.

Generalizing eq.~\m, by introducing an extra
undetermined mass parameter $m_{i0}$ for each
flavour, we define $\rho_i ={m_i/m_{i0}}$.
The mass term has the form
$$M.T.=C\tilde m N_{\tilde m}
\Tr\left(M\,g\right)=C\sum_{i=1}^{N_f}m_{i0}\tilde m N_{\tilde m}
\rho_i g_{ii}\eqn\amiVI$$
By integrating out the gauge fields and taking the strong-coupling
limit, we get
$$\tilde S\left\( g\right\)=N_cS\left\( g\right\)+\sum_{i=1}^{N_f}
\hat m_i^2N_{\hat m_i}\int d^2x{m_i\over m_{i0}}(g_{ii}+g^*_{ii})
 \eqn{\eqscI}$$
$$\hat m_i
=\left\( C N_cm_{i0}\left(e_c^2N_f\over 2\pi\right)^{{1\over 2}p}
\right\)^{1\over p+1} \eqn\amiVI$$
The classical low mass solutions to the action \eqscI\ are given by
\gx\ with an action density \sdg, but now with mass parameters
$$\tilde m_i^2=\rho_i \hat m_i^2=
{m_i\over m_{i0}}\hat m_i^2\eqn\amiVII$$
Each $\phi_i$ has a classical solution \fix\ with energy \ei.
The minimum classical energy solution is given by the ansatz \g0, with
$\tilde m_{N_f}$ chosen such that it is the smallest mass.
Define now $A(t)$ as in \At, resulting in an effective action like in
\dsg, and  a Hamiltonian
$$H=M\left\{1+\left({\pi\over 2N_c}\right)^2
 \left\(C_2(R)-{N_c^2\over 2N_f} (N_f-1)\right\)+\sum_{i=1}^{N_f}
\left\({m_i\over m_{N_f}}\left({m_{i0}\over
m_{N0}}\right)^{1-p\over 1+p }-1
\right\)\vert z_i\vert^2\right\}\eqn\amiVIII$$
$$M=4\tilde m_{N_f}\left({2N_c\over\pi}\right)^{1\over 2}.$$
There are several problems connected with these arbitrary mass
parameters.
Since the choice of the parameters $m_{i0}$ is arbitrary, we can choose
$$m_{i0}=m_{N0}\left({m_i\over m_{N_f}}\right)^{p+1\over p-1},$$
in which case
$H$ has no explicit contribution due to the mass differences,
which is unacceptable.

The question now arises at what stage should one apply the
semi-classical approximation.\refmark{\FZ}
The classical solution is governed by $\tilde m_i$ but the lightest
$\tilde m_i$ is not necessarily the lightest $m_i$.
In fact,
the normal ordering ambiguity is present even in a
non-interacting theory. To see this, we formally set
$N_c=1$, i.e.  we have the field $g$ with level 1 which
means free quarks. There is no gauge group, so the dependence on $e_c$
should vanish and indeed $p=0$, $\hat m_i=Cm_{i0}$ and
\hbox{$M=4C\left({\ds 2m_{N_f}m_{N0}\over\ds \pi}\right)^{1\over 2}$.}
The lowest multiplet
corresponds to a Young tableaux of one box and
\hbox{
$C_2(R)=\ds {\ds N_f^2-1\over \ds \phantom{^a}2N_f\phantom{^a}}$,
}
therefore
$$ H=M\left\(\left({\pi\over 2}\right)^2{N_f-1\over 2}+\sum_{i=1}^{N_f}
{m_im_{i0}\over m_{N_f}m_{N0}}\vert z_i\vert^2\right\)
\eqn\amiIX$$
which in general does not look like a non-interacting theory,
unless one takes $m_{i0}=m_{N0}$ for all $i$.

One could argue that the above example is not relevant
for a generic case, since for small $N_c$
the semi-classical approximation is not justified.
This can be seen from \H,
as follows. Consistency with the semi-classical
approximation requires the
quantum corrections to the energy to be much smaller than the classical
contribution. In the case of one box this means
$\left({\ds \pi\over \ds 2}\right)^2{\ds {N_f}_{}-1
\over\phantom{^\dagger} \ds 2\phantom{\dagger}}\ll 1$,
which for any $N_f\geq2$ is not correct.

Nevertheless, since the choice of the $m_{i0}$'s is {\sl a priori}
arbitrary,
we see that there is an ambiguity. In the non-interacting case
it is obvious what is the right choice of the $m_{i0}$'s,
but in general we do not know what is the choice
leading to  the optimal
semi-classical approximation, as is best illustrated by
the case of equal quark masses, which can be made to appear
unequal by a suitable choice of the $m_{i0}$'s.

One can summarize the situation as follows.
When all masses are equal, the treatment of ref.~[\DFS],
in the strong coupling limit, yields an effective Lagrangian
in terms of flavour degrees of freedom only, with
a mass scale that involves the coupling $\alpha_c$, as in
eq.~\amiII.
When the current mass $M$ of some quark is heavy,
$M^2\gg\alpha_c$, we expect the constituent mass of that quark to
be heavy and close to the current mass.
The problems arise in the intermediate cases, when we do
not know what is the ``best" starting point for the normal ordering
scale, before going to the semi-classical approximation.
Of course, if one were able to sum all corrections, one would obtain
the full result, regardless of the starting point.

\refout
\endpage

\def\nl{\nextline}

\FIG\FigI{Mesons in QED$_2$ with two flavours, solutions
of eqs.~\QEDIV, \QEDXI.\nl
(a) $\alpha =0.1$, $m_1=0.5$, $m_2=0.5$;\nextline
(b) $\alpha =0.1$, $m_1=2.0$, $m_2=0.5$;\nextline
(c) $\alpha =0.1$, $m_1=4.0$, $m_2=0.5$;\nextline
(d) $\alpha =2.0$, $m_1=4.0$, $m_2=0.5$.\nextline
In all four plots the upper line corresponds to $\chi_1$,
and the lower line to $\chi_2$. (a), (b) and (c) show
how, with increasing $m_1$, the $\chi_1$ profile converges
to a step function, while the light flavour profile, $\chi_2$,
remains almost unaffected.
The $\alpha $ dependence can be inferred from the
difference between (c) and (d): $\chi_2$ gets heavier with
increasing $\alpha/m_2^2$. The effect is not a sharp one,
because part of the $\alpha$ dependence
is already included in the value of $m_2$, through
the normal ordering (see Appendix).}

\FIG\FigII{Mesons in QCD$_2$ with $N_c=2$, $N_f=2$, solutions
of eqs.~\VII, \XV.\nl
(a) $\ac =1.0$, $m_1=0.1$, $m_2=0.1$;\nextline
(b) $\ac =1.0$, $m_1=0.1$, $m_2=0.5$;\nextline
(c) $\ac =1.0$, $m_1=0.1$, $m_2=2.0$;\nextline
(d) $\ac =10.0$, $m_1=0.1$, $m_2=2.0$.\nextline
$\chi_{11}$ and $\chi_{12}$ are the upper and lower continuous
curves, respectively. The ``non-valence" components
$\chi_{21}$ and $\chi_{22}$ are denoted by dot-dashed and dashed
lines, respectively, except for (a), where they are exactly
zero.}

\FIG\FigIII{Baryons in QCD$_2$ with $N_c{=}3$, $N_f{=}3$, solutions
of eq.~\VII.\nl
(a) a ``$uuu$"-like baryon, eq.~\FrozenColour, with
$\ac {=}1.0$, $m_1{=}m_2{=}m_3{=}0.1$;\nl
(b) a ``$ucb$"-like baryon, eq.~\genuineSolution, with
$\ac {=}1.0$, $m_1{=}0.1$, $m_2{=}0.5$, $m_3{=}0.8$;\nl
(c) a ``$ub\,t$" -like baryon, eq.~\genuineSolution, with
$\ac {=}1.0$, $m_1{=}0.1$, $m_2{=}1.0$, $m_3{=}2$.\nextline
In (a) the ``non-valence" components, $\chi_{21}$, $\chi_{31}$, etc.
are exactly zero.
In (b) and (c) the continuous lines denote the ``valence"
components, while the dashed and dot-dashed lines denote the
``non-valence" components. The ``non-valence" components
of the heavy flavours are clearly much more suppressed than
those of the light flavour.}

\baselineskip=18pt
\figout
\bye